\documentclass[a4paper,11pt]{article}

\usepackage{amsmath}
\usepackage{pos}

\newlength{\bibitemsep}
\newlength{\bibparskip}
\let\oldthebibliography\thebibliography
\renewcommand\thebibliography[1]{%
	\oldthebibliography{#1}%
	\setlength{\parskip}{\bibitemsep}%
	\setlength{\itemsep}{\bibparskip}%
}

\title{Direction reconstruction for the in-ice radio array of IceCube-Gen2}

\ShortTitle{Direction reconstruction for the in-ice radio array of IceCube-Gen2}

\author{The IceCube-Gen2 Collaboration \\{\normalsize \normalfont(a complete list of authors can be found at the end of the proceedings)}\\}

\emailAdd{sjoerd.bouma@fau.de}
\emailAdd{anna.nelles@desy.de}

\abstract{

The IceCube-Gen2 facility will extend the energy range of IceCube to ultra-high energies. The key component to detect neutrinos with energies above 10 PeV is a large array of in-ice radio detectors. In previous work, direction reconstruction algorithms using the forward-folding technique have been developed for both shallow ($\lesssim 20$ m) and deep in-ice detectors, and have also been successfully used to reconstruct cosmic rays with ARIANNA. Here, we focus on the reconstruction algorithm for the deep in-ice detector, which was recently introduced in the context of the Radio Neutrino Observatory in Greenland (RNO-G).

We discuss the performance-critical aspects of the algorithm, as well as recent and future improvements, and apply it to study the performance of a station of the IceCube-Gen2 in-ice radio array. We obtain the angular resolution, which turns out to be strongly asymmetric, and use this to optimize the configuration of a single station.

\vspace{4mm}
{\bfseries Corresponding authors:}
Sjoerd Bouma$^{1}$, Anna Nelles$^{1,2*}$\\
{$^{1}$ \itshape Erlangen Centre for Astroparticle Physics, Friedrich-Alexander-Universit{\"a}t Erlangen-N{\"u}rnberg, D-91058 Erlangen, Germany}\\
{$^{2}$ \itshape Deutsches Elektronen-Synchrotron DESY, Platanenallee 6, 15738 Zeuthen, Germany}\\[4mm]
$^*$ Presenter

\ConferenceLogo{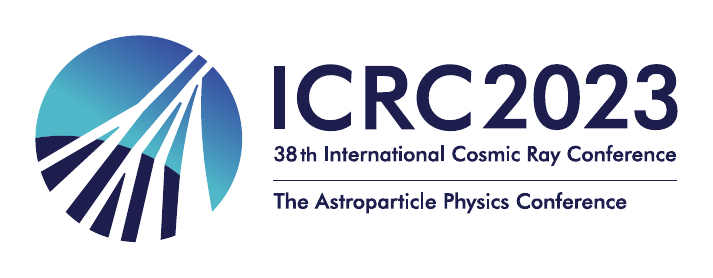}

\FullConference{The 38th International Cosmic Ray Conference (ICRC2023)\\ 26 July -- 3 August, 2023\\ Nagoya, Japan}
}

\begin{document}

\maketitle

\section{Introduction}\label{sec:introduction}
The in-ice radio array of IceCube-Gen2 will extend the energy range of the facility to ultra-high energies (UHE) above 10 PeV by exploiting the radio emission of in-ice particle showers due to the Askaryan effect. The in-ice radio array is planned to consist of both hybrid detector stations, with both shallow and deep ($\sim150$ m) antennas, as well as shallow-only stations \cite{TDR}. At completion, the array will cover about 500 km$^2$ and improve on current detectors by an order of magnitude in sensitivity, providing the best chance yet of measuring the flux at this previously inaccessible energy range, as well as establishing a direct connection between the highest-energy cosmic rays and neutrinos.
To improve the likelihood of their cosmic origin, and enable neutrino astronomy at these energies, the ability to reconstruct the direction of triggering neutrino events is crucial. 

\section{The reconstruction algorithm}\label{sec:algorithm}
The reconstruction algorithm used in these proceedings is the forward-folding algorithm \cite{Glaser:2019rxw,Barwick:2021wU,Arianna:2021lnr,Plaisier:2023cxz}. The forward-folding technique has been successfully used to reconstruct cosmic rays with the shallow upward-facing antennas  \cite{Arianna:2021lnr} and has been applied to the shallow detector component of IceCube-Gen2 finding an average resolution of the neutrino direction of $\sigma_{68\%}= 3^\circ$ \cite{Barwick:2021wU}. As most events that are detected by the deep component are not expected to be visible in the shallow antennas, an improved algorithm for these events has been developed in \cite{Plaisier:2023cxz} and applied to an RNO-G like detector in Greenland. Here, we apply this algorithm to study the expected performance of the deep component of a single station of IceCube-Gen2. The algorithm is available as part of the open-source NuRadioMC framework \cite{Glaser:2019cws}. In this section, we only provide a brief summary of this algorithm; for more details, we refer to \cite{Plaisier:2023cxz}.

\begin{figure}
	\centering
	\includegraphics[width=.95\textwidth, trim={0 45 0 0}, clip=true]{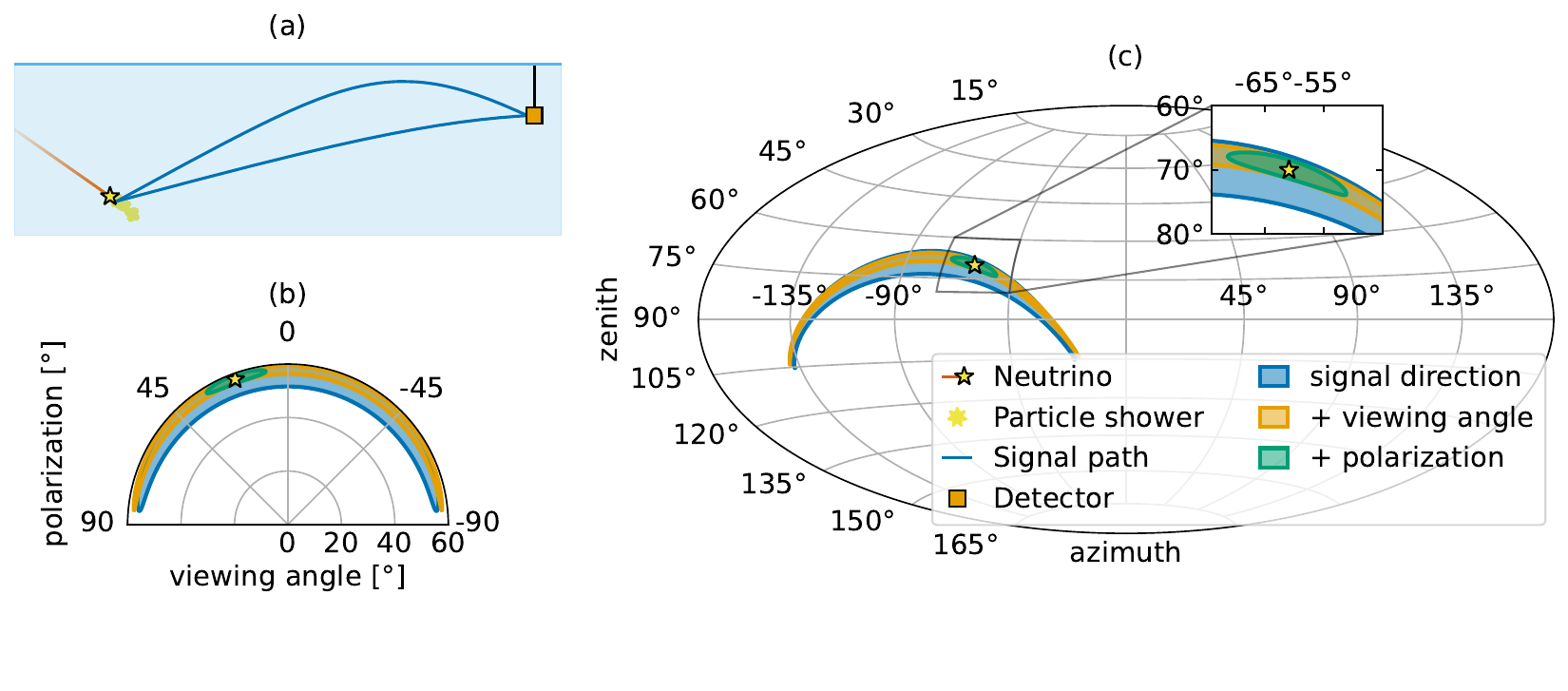}
	\caption{Illustration of the direction reconstruction algorithm. (a) The neutrino interacts in the ice, producing a particle shower and Askaryan emission, which travels along curved trajectories in the ice to reach the detector. (b) The signal direction (at (0,0)) restricts the possible neutrino directions to a broad cone, which is defined by the viewing angle ($\approx$ Cherenkov angle) and the polarization. Their determination then allows to uniquely identify the original neutrino direction. (c) The same plot as (b), in local on-sky coordinates.}
	\label{fig:onsky_didactic}
\end{figure}

The reconstruction relies on two features of the Askaryan effect that causes the radio emission which is detected. Firstly, the shape of the frequency spectrum depends on the \textbf{viewing angle} under which the signal is observed at the detector. The closer this is to the Cherenkov angle ($\approx 56^\circ$ in deep polar ice), the larger the contribution of higher frequency components. Secondly, the polarization vector points towards the shower axis. Thus, the combined determination of the \textbf{signal emission direction}, \textbf{viewing angle} and \textbf{polarization} is sufficient to uniquely determine the direction of the original neutrino. This is illustrated in Fig.~\ref{fig:onsky_didactic}.

These requirements inform the design of a deep in-ice detector. As most events are expected to be detected only in a single detector station, each of these consists of three vertical 'strings' in a triangular configuration, to enable the determination of the emission vertex. These strings are populated with omnidirectional antennas that are sensitive either to vertically polarized ('Vpol') or horizontally polarized ('Hpol') signals at various depths, which provide sensitivity to the signal polarization. A sketch of the reference design for the deep component for IceCube-Gen2 is shown in Fig.~\ref{fig:detector_layouts}.

In order to limit the computation time required to reconstruct a single event, the current approach is to split the reconstruction into two steps. The first step is to reconstruct the position of the shower maximum, which is treated as the origin of the radio signal using a ray-tracing approximation. The algorithm used for this step is heavily based on the one outlined in \cite{Aguilar:2021uzt}, referred to therein as vertex reconstruction. Note that, however, the position of the shower maximum is not the same as the neutrino interaction vertex; the difference is of the order of 10 m for hadronic showers, and up to an order of magnitude larger for high-energy electromagnetic showers subject to the LPM effect.
An additional complication arises due to the fact that the index of refraction in polar ice is not constant. This results in multiple curved and possibly reflected ray trajectories, referred to as 'ray types'. The obtained signal direction therefore depends also on the ice model and the 'ray type'. For the former, we use the simple exponential SPICE 2015 model described in \cite{Barwick_2018} both in simulation and reconstruction; the impact of systematic uncertainties and features such as layers or birefringence is beyond the scope of this work. The dominant ray type is subsequently identified by template correlation.

The second part of the reconstruction is the determination of the viewing angle and polarization. This is done using a 'forward-folding' approach: the modelled Askaryan signal is propagated through the ice and convolved with the detector response, after which the $\chi^2$ statistic is computed and minimized between the measured and modelled voltage traces. For in-ice radio detectors, this approach was shown to offer superior performance in the reconstruction of cosmic rays, particularly at low signal-to-noise ratios (SNR), compared to traditional 'unfolding' methods \cite{Glaser:2019rxw}.

With respect to the algorithm described in \cite{Plaisier:2023cxz}, a small number of changes and improvements has been made:
\begin{enumerate}
	\item A minimizer, which is initialized at the most common degenerate solutions, has been implemented in addition to the existing brute-force optimization. In $\sim 90 \%$ of cases, the same (global) minimum is found as previously, but the computation time is reduced by more than an order of magnitude, enabling (relatively) efficient studies to e.g.\ compare the performance of different detector layouts, as done here.
	\item Due to imperfect vertex reconstruction and, in more realistic scenarios, slight mismodelling of the detector geometry or the ice, the exact pulse arrival time at each antenna has to be determined by an additional correlation step. For very low SNR pulses, this will select and fit a random noise fluctuation, resulting in an overestimation of the actual signal contribution in this antenna. This is prevented in two ways; antennas below a certain SNR threshold are excluded from the fit, and for Hpol antennas (which generally have much less signal than the Vpol antennas), the timing is determined from an adjacent Vpol antenna that does pass the threshold. This latter strategy was previously used only for the Hpol antennas on the central string, but has now also been implemented for the secondary strings.
	\item Finally, the algorithm in \cite{Plaisier:2023cxz} was designed to reconstruct hadronic showers only. As in reality a mix of hadronic and electromagnetic showers is expected, a fit for electromagnetic showers has been implemented, and the event is tagged based on which hypothesis results in the overall minimum $\chi^2$. For low to medium quality events, this tagging is fairly inaccurate, but does not noticeably deteriorate performance, whereas the resolution for high-quality electromagnetic-dominated events is improved.
\end{enumerate}

\section{Performance}\label{sec:performance}
The performance of the algorithm is evaluated by simulating neutrinos at half-decade intervals ranging from $E_\nu=10^{16.5}-10^{19.0}$ eV using NuRadioMC \cite{Glaser:2019cws}. We include both hadronic and electromagnetic (due to $\nu_e$ charged-current interactions) events, assuming a 1:1:1 flavour ratio at the detector. The semi-analytical model described in \cite{Alvarez-Muniz:2011wcg}, which includes different stochastic realizations of the shower profile, is used for the simulation. The reconstruction is performed using a fully analytic model \cite{Alvarez-Muniz:2010hbb}. The detector layout used is the 'reference' design \cite{TDR} for an IceCube-Gen2 hybrid detector, with a 4-channel phased-array at 150 m depth used for triggering at a target trigger rate of 100 Hz.

\begin{figure}
	\centering
	\includegraphics[width=.88\textwidth]{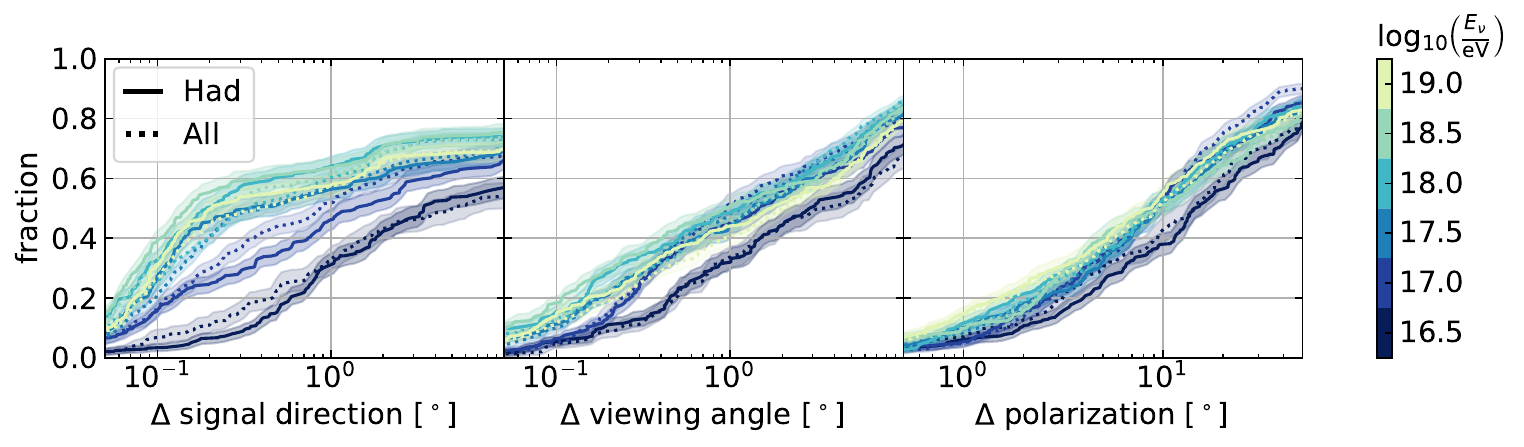}
	\caption{Performance of the reconstruction algorithm, shown as the cumulative distribution of the difference between the true and reconstructed quantity, versus neutrino energy $E_\nu$. Note the different horizontal scale on the third plot. Shaded regions indicate the $1\sigma$ statistical uncertainty due to the limited sample size.}
	\label{fig:1d_cdfs_vE}
\end{figure}

The results of the reconstruction are 
shown separately for each aspect of the reconstruction in Fig.~\ref{fig:1d_cdfs_vE}. We distinguish different simulated neutrino energies, and additionally show results both for hadronic events only ('Had') and those for the overall mix of hadronic and hadronic+electromagnetic events ('All').

Particularly at low energies, the overall reconstruction is limited by the performance of the shower maximum reconstruction, which determines the signal direction. These relatively low-energy events are generally lower in SNR, and originate from closer to the detector and nearer to the ice surface. They are more challenging to reconstruct in general, but especially for the shower maximum reconstruction, which requires a signal in all three strings and at least two different depths to be unambiguous. If either the position of the shower maximum or the appropriate ray type is not determined successfully, this may additionally impact the subsequent aspects of the reconstruction by assuming the wrong pulse position within the voltage traces.

If the shower maximum reconstruction is successful, the overall error is instead dominated by the error on the polarization angle. This is due both to the much larger phase space for this quantity compared to the viewing angle, which is constrained to lie within a couple of degrees of the Cherenkov angle, and the fact that the Hpol antennas are less sensitive and generally record a much smaller contribution than the Vpol antennas.

In general, the reconstruction performs best for the 'middle' energies. At the lowest energies, the lower SNR and more challenging shower maximum reconstruction strongly limit the fraction of triggered events that can be reconstructed well. At the highest energies, as the shower shapes become more irregular, performance once again starts to decrease. Finally, the performance for electromagnetic events at low energies is slightly better than that for purely hadronic events. This can be partially explained by considering that for electromagnetic events, the entire neutrino energy is deposited in the particle showers, leading to a larger signal amplitude at the detector.



\begin{figure}
	\centering
	\includegraphics[width=.95\textwidth]{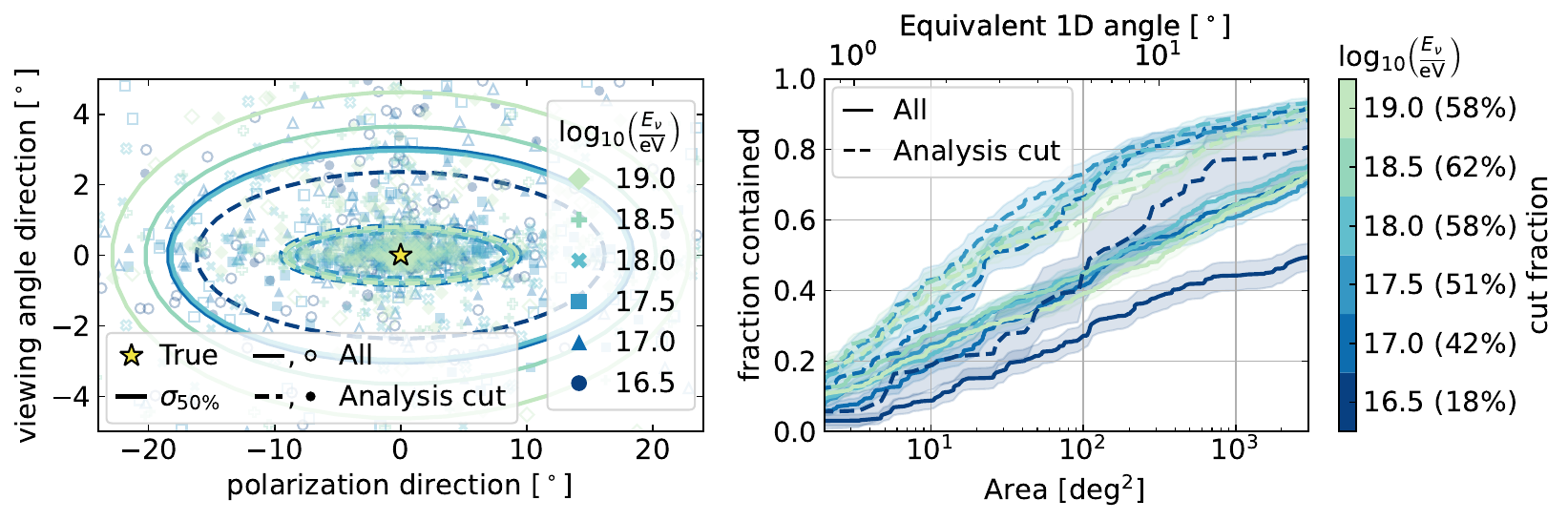}
	\caption{
		The 2D distribution of all reconstructed events. Left: After rotation and projection to the polarization and viewing angle directions, the (reconstructed $-$ true) neutrino directions are distributed along elongated ellipses. Shown are individual reconstructed events, as well as the ellipse containing 50\% of all events. 
		Events passing the analysis cut are indicated with filled symbols and dashed ellipses. 
		Right: the fraction of reconstructed events contained within an ellipse of a certain area, for both all events and those passing the analysis cut. Note that this does \textbf{not} imply that these events also have single-event N\% uncertainty contours that are smaller than the given ellipse. The top x-axis shows the area-equivalent 1D angle for a symmetric contour.}
	\label{fig:2d_cdfs_vE}
\end{figure}

Because the overall error is dominated by the polarization, we do not quote the space angle difference. Instead, we proceed as follows: we project the reconstructed neutrino direction onto a frame where the polarization direction lies along the x-axis, and the viewing angle direction along the y-axis. This is shown in Fig.~\ref{fig:2d_cdfs_vE}, left. We then quote the fraction of reconstructed events contained within the area of an elongated ellipse centred on the true neutrino direction. This gives a more accurate impression of the actual 2d spread of the reconstruction algorithm; however, it should be stressed that the area within which a single event reconstructs does \textbf{not} allow to make inferences on the true single-event uncertainty, which may be smaller or larger. A procedure to determine single-event uncertainty contours using resimulation has been described in \cite{Plaisier:2023cxz}; as this process takes of the order of several hours per event, we do not use this here. 

Instead, we show how the resolution can be improved by applying an analysis cut, which is here defined by requiring a minimum SNR of 2.5 in two antennas on the central string, of which at least one should be at a different depth than the phased array, as well as an SNR $>2$ on at least one antenna in both of the secondary strings. The fraction of events retained at each energy after this cut is shown on the right of Fig.~\ref{fig:2d_cdfs_vE}. Note that this cut has been defined based on the true (simulated) signal SNR in the [96, 500] MHz band; it is expected that a similar cut can be defined based on data with a moderate loss of either efficiency or resolution.

\section{Comparison of different layouts}\label{sec:comparison}
In addition to evaluating the performance of the 'reference' design of the IceCube-Gen2 hybrid radio station, we have also performed the reconstruction for several alternative detector layouts. 
In order to be able to compare the performance of different layouts consistently, the simulation was only performed once, using an 'overinstrumented' detector containing all antennas one wants to consider. As the trigger remains the same for all layouts considered here, the same events can be used for the comparison by simply changing the antennas used in the reconstruction.

\begin{figure}
	\centering
	\includegraphics[width=.77\textwidth]{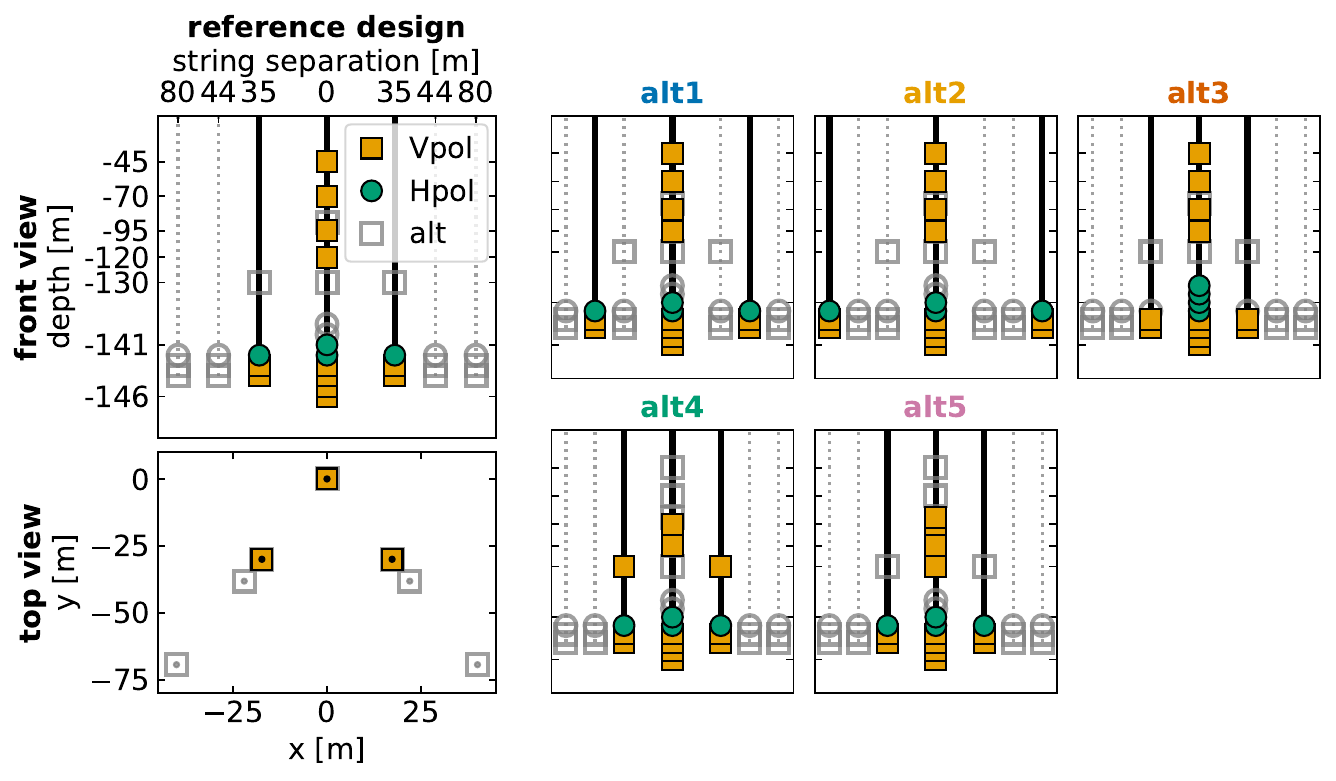}
	\caption{Different detector layouts used for the performance comparison. 
		The central string contains the four-channel phased array used for triggering at the bottom. The antennas present in each configuration are shown in orange and green, with the unused antennas in grey.}
	\label{fig:detector_layouts}
\end{figure}

A sketch of all layouts used in this study is shown in Fig.~\ref{fig:detector_layouts}. 
All designs include 16 deep in-ice antennas, leaving 8 of the total 24 DAQ channels for the shallow component.
In addition to the reference design, we further consider:
\begin{itemize}
	\item The \textbf{alt1} and \textbf{alt2} designs increase the horizontal baseline of the three strings. A larger baseline helps to more accurately constrain the position of the shower maximum, and may also help to constrain the polarization by mapping out the Cherenkov cone, at the cost of the loss of SNR as one goes away further from the triggering antenna.
	\item The \textbf{alt3} layout moves all four Hpol antennas to the central string. In general, the highest SNR is expected close to the triggering antennas; having all four Hpol antennas directly above the phased array may therefore help to increase the signal contribution in these antennas.
	\item Finally, the \textbf{alt4} and \textbf{alt5} designs move the two shallower Vpol antennas deeper. A signal in one of the upper Vpol antennas can provide a strong additional constraint in the reconstruction, particularly for the position of the shower maximum. However, due to their relatively large separation from the triggering channels, as well as the 'shadow-zone' \cite{Barwick_2018} due to the downward-bending of in-ice radio signals, detectable signals in these antennas are rare, motivating these designs with smaller vertical baselines, which are expected to contribute for a larger fraction of triggered events.
\end{itemize}

\begin{figure}
	\centering
	\includegraphics[width=.90\textwidth]{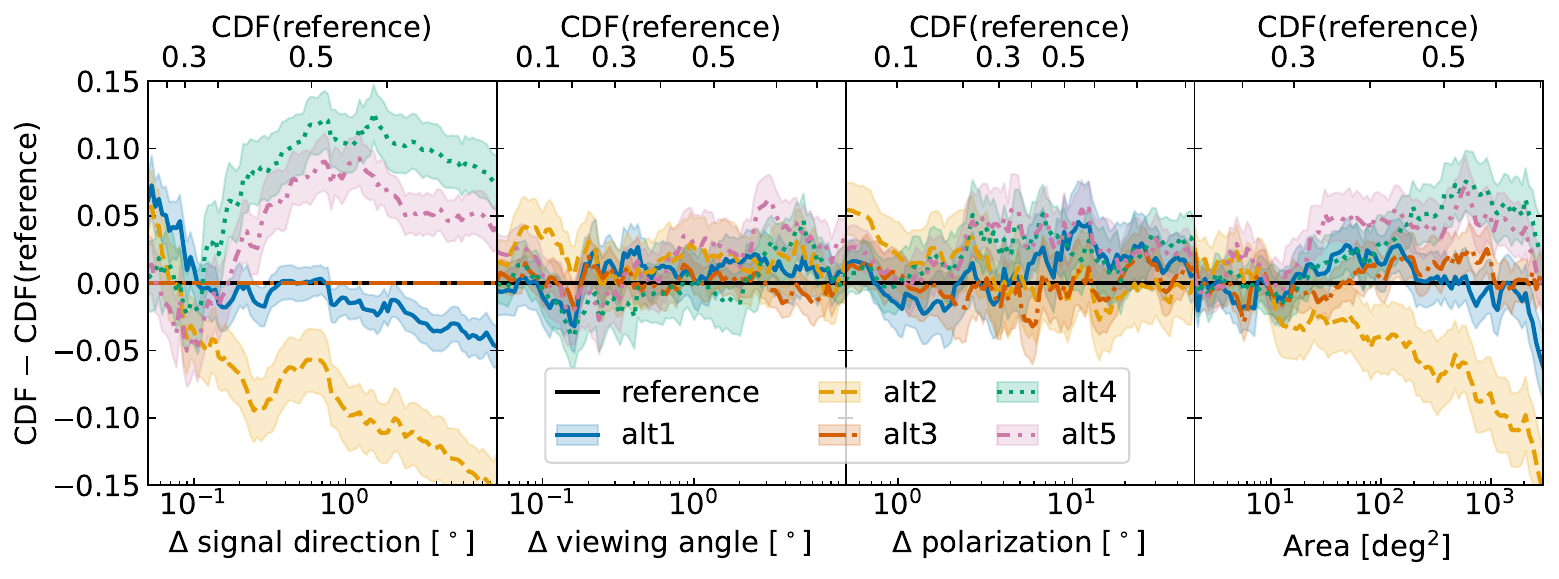}
	\caption{Comparison of the performance of the different detector layouts. The improvement is shown as the additional fraction of all events that reconstructs within a certain $\Delta$, compared to the reference layout, for which the CDF is shown on the top x-axis.}
	\label{fig:comparison}
\end{figure}

The results for each configuration are shown in Fig.~\ref{fig:comparison}. They show the improvement in performance as the additional fraction of events that reconstructed within a certain resolution, i.e.\ the difference between the respective CDFs. We compare the performance across all energies by reweighting with the expected flux at the detector, assuming a single power-law spectrum \cite{IceCube_flux} plus a GZK component from cosmogenic neutrinos \cite{GZK}. Only hadronic events were included for the comparison, although as only the performance relative to the reference is shown here, the impact of excluding electromagnetic events here can be assumed to be small.

We make the following observations: Increasing the horizontal baseline (configurations alt1, alt2) mainly improves the precision of the signal direction determination, but only for the $\sim30\%$ best events. The error for these events, however, is dominated by the uncertainties on viewing angle and polarization, where no clear improvement is observed. For lower quality events, on the other hand, the lower signal contribution in the now further-away secondary strings hinders the reconstruction of the shower maximum, especially for the largest baseline (alt2).

Moving all four Hpol antennas to the central string (alt3) does not affect the signal direction reconstruction, which relies only on the Vpol antennas. No clear improvement is shown for the viewing angle or polarization reconstruction, either, however. (Note that, before improvement 2 in Sec.~\ref{sec:algorithm} had been implemented, this configuration did show a $\sim5\%$ improvement in the number of events reconstructed with a polarization error of $\le 10^\circ$).

Finally, replacing the two uppermost Vpol antennas with deeper antennas (alt4, alt5) significantly increases the fraction of events with a signal direction reconstructed within O(1) deg, without negatively impacting the resolution obtained by the other parts of the algorithm. Most of the improvement, however, seems to apply to low or medium quality events, such that the increase in the fraction of events reconstructing within a certain area is relatively modest.

In general, the impact of changing the detector configuration is not dramatic, and larger improvements are expected by further refinements of the reconstruction algorithm, or the use of machine-learning based approaches such as \cite{Glaser:2022lky, Heyer:2023}.

\section{Conclusion}\label{sec:conclusion}
We have applied the forward-folding reconstruction algorithm outlined in \cite{Plaisier:2023cxz} to the reference design for the deep component of the in-ice radio detector of IceCube-Gen2. Although the detector, ice and exact trigger configuration are different, the overall resolution obtained is similar to that for the RNO-G-like station in \cite{Plaisier:2023cxz} for purely hadronic events, and slightly improved for events with electromagnetic showers, thanks to the addition of a new step in the fitting algorithm. As in that work, the resolution obtained is much larger in one direction than in the other, necessitating a full 2D treatment in order to accurately capture the actual uncertainty.
In the second part, we have compared the performance of different layouts of the deep component. We conclude that, in particular for the viewing angle and the polarization, the exact positioning of the additional Vpol antennas only has a minor impact on the reconstruction. Conversely, the reconstruction of the signal direction is shown to improve for designs which move some of the shallower Vpol antennas closer to the bottom of the station, which on average increases the number of channels with a detectable signal contribution. 

\bibliographystyle{ICRC}
\bibliography{references}

%

\clearpage

\section*{Full Author List: IceCube-Gen2 Collaboration}

\scriptsize
\noindent
R. Abbasi$^{17}$,
M. Ackermann$^{76}$,
J. Adams$^{22}$,
S. K. Agarwalla$^{47,\: 77}$,
J. A. Aguilar$^{12}$,
M. Ahlers$^{26}$,
J.M. Alameddine$^{27}$,
N. M. Amin$^{53}$,
K. Andeen$^{50}$,
G. Anton$^{30}$
% ...
\\
\\
% ...
$^{12}$ Universit{\'e} Libre de Bruxelles, Science Faculty CP230, B-1050 Brussels, Belgium \\
% ...
$^{17}$ Department of Physics, Loyola University Chicago, Chicago, IL 60660, USA \\
% ...
$^{22}$ Dept. of Physics and Astronomy, University of Canterbury, Private Bag 4800, Christchurch, New Zealand \\
% ...
$^{26}$ Niels Bohr Institute, University of Copenhagen, DK-2100 Copenhagen, Denmark \\
$^{27}$ Dept. of Physics, TU Dortmund University, D-44221 Dortmund, Germany \\
% ...
$^{30}$ Erlangen Centre for Astroparticle Physics, Friedrich-Alexander-Universit{\"a}t Erlangen-N{\"u}rnberg, D-91058 Erlangen, Germany \\
% ...
$^{47}$ Dept. of Physics and Wisconsin IceCube Particle Astrophysics Center, University of Wisconsin{\textendash}Madison, Madison, WI 53706, USA \\
% ...
$^{50}$ Department of Physics, Marquette University, Milwaukee, WI, 53201, USA \\
% ...
$^{53}$ Bartol Research Institute and Dept. of Physics and Astronomy, University of Delaware, Newark, DE 19716, USA \\
% ...
$^{76}$ Deutsches Elektronen-Synchrotron DESY, Platanenallee 6, 15738 Zeuthen, Germany  \\
$^{77}$ Institute of Physics, Sachivalaya Marg, Sainik School Post, Bhubaneswar 751005, India \\
% ...

\subsection*{Acknowledgements}

\noindent
The authors gratefully acknowledge the support from the following agencies and institutions:
USA {\textendash} U.S. National Science Foundation-Office of Polar Programs,
U.S. National Science Foundation-Physics Division,
U.S. National Science Foundation-EPSCoR,
Wisconsin Alumni Research Foundation,
Center for High Throughput Computing (CHTC) at the University of Wisconsin{\textendash}Madison,
Open Science Grid (OSG),
Advanced Cyberinfrastructure Coordination Ecosystem: Services {\&} Support (ACCESS),
Frontera computing project at the Texas Advanced Computing Center,
U.S. Department of Energy-National Energy Research Scientific Computing Center,
Particle astrophysics research computing center at the University of Maryland,
Institute for Cyber-Enabled Research at Michigan State University,
and Astroparticle physics computational facility at Marquette University;
Belgium {\textendash} Funds for Scientific Research (FRS-FNRS and FWO),
FWO Odysseus and Big Science programmes,
and Belgian Federal Science Policy Office (Belspo);
Germany {\textendash} Bundesministerium f{\"u}r Bildung und Forschung (BMBF),
Deutsche Forschungsgemeinschaft (DFG),
Helmholtz Alliance for Astroparticle Physics (HAP),
Initiative and Networking Fund of the Helmholtz Association,
Deutsches Elektronen Synchrotron (DESY),
and High Performance Computing cluster of the RWTH Aachen;
Sweden {\textendash} Swedish Research Council,
Swedish Polar Research Secretariat,
Swedish National Infrastructure for Computing (SNIC),
and Knut and Alice Wallenberg Foundation;
European Union {\textendash} EGI Advanced Computing for research;
Australia {\textendash} Australian Research Council;
Canada {\textendash} Natural Sciences and Engineering Research Council of Canada,
Calcul Qu{\'e}bec, Compute Ontario, Canada Foundation for Innovation, WestGrid, and Compute Canada;
Denmark {\textendash} Villum Fonden, Carlsberg Foundation, and European Commission;
New Zealand {\textendash} Marsden Fund;
Japan {\textendash} Japan Society for Promotion of Science (JSPS)
and Institute for Global Prominent Research (IGPR) of Chiba University;
Korea {\textendash} National Research Foundation of Korea (NRF);
Switzerland {\textendash} Swiss National Science Foundation (SNSF);
United Kingdom {\textendash} Department of Physics, University of Oxford.

\end{document}